\documentclass[a4paper,11pt]{article}

\usepackage{amssymb}
\usepackage{amsmath}
\usepackage{bbm}
\usepackage{amscd}
\usepackage{latexsym}
\usepackage{epsfig
}
\usepackage{color}
\usepackage{graphicx}
\usepackage{booktabs}
\usepackage{a4wide}
\usepackage[bf,footnotesize]{caption}

\usepackage{graphicx} 
\usepackage{tikz} 
\usetikzlibrary{arrows,positioning,automata,shadows,fit,shapes} 

\usepackage[colorlinks=true,a4paper=true,backref=false,linktocpage=true,citecolor=blue,urlcolor=blue,linkcolor=blue,pdfpagemode=UseOutlines]{hyperref}

\hypersetup{%
  bookmarksnumbered=true,
  pdftitle = {},
  pdfsubject = {},
  pdfauthor = {U.~Wenger},
  pdfkeywords = {}
}

\DeclareMathOperator{\Tr}{Tr}

\title{Loop formulation of supersymmetric Yang-Mills quantum mechanics}

\author{Kyle Steinhauer and Urs Wenger \vspace{0.5cm}
\\
        Albert Einstein Center for Fundamental Physics\\
         Institute for Theoretical Physics\\
        University of Bern\\
        Sidlerstrasse 5\\
        CH--3012 Bern\\
        Switzerland\vspace{0.5cm}\\}

\date{\empty}

\begin{document}

\maketitle

\begin{abstract}
\noindent
We derive the fermion loop formulation of ${\cal N}=4$ supersymmetric
SU($N$) Yang-Mills quantum mechanics on the lattice. The loop
formulation naturally separates the contributions to the partition
function into its bosonic and fermionic parts with fixed fermion
number and provides a way to control potential fermion sign problems
arising in numerical simulations of the theory. Furthermore, we
present a reduced fermion matrix determinant which allows the
projection into the canonical sectors of the theory and hence
constitutes an alternative approach to simulate the theory on the
lattice.
\end{abstract}

\section{Introduction}
The conjectured holographic duality between supersymmetric Yang-Mills
quantum mechanics and the theory of D0 branes of type IIa string
theory in the large-$N$ limit in principle allows to probe the physics
of certain supergravity black holes by lattice Monte Carlo
simulations. In particular, ${\cal N}=16$ supersymmetric Yang-Mills
(SYM) quantum mechanics (QM) stemming from the dimensional reduction
of ${\cal N}=1$ SYM in $d=10$ dimensions is supposed to describe the
dynamics of D0 branes which are the degrees of freedom of the
underlying M-theory \cite{Banks:1996vh}. The connection to so-called
black $p$-branes allows to study the thermodynamics of black holes
through the corresponding strongly coupled gauge theory. We refer the
reader to the review article \cite{Taylor:2001vb} for further
details. Here we report on our work in this direction on an analogue,
but simpler theory, namely ${\cal N}=4$ SYM QM with generic gauge
group SU($N$). The model stems from dimensionally reducing ${\cal
  N}=1$ SYM in $d=4$ dimensions, but is expected to share many
qualitative features with the 16 supercharge model. The aim of this
paper is to construct the fermion loop formulation of the strongly
coupled gauge theory regularised on the lattice, so as to make it
susceptible to numerical simulations.

There have already been a number of nonperturbative investigations of
SYM QM using numerical techniques. In \cite{Wosiek:2002nm,
  Campostrini:2004bs, Veneziano:2005qs, Veneziano:2006bx,
  Veneziano:2006cx} the Hamiltonian formulation was employed together
with the cut Fock space method. This approach also allowed analytic
solutions, at least for $d=2$ dimensional SYM QM
\cite{Campostrini:2002mr, Korcyl:2011tq, Korcyl:2011vr, Korcyl:2010uq,
  Korcyl:2009yh}. On the other hand, in \cite{Janik:2000tq} the Wilson
lattice discretization was constructed and the $d=4$ SYM QM was
simulated in the quenched approximation \cite{Janik:2000tq,
  Kawahara:2007fn}. Further discretizations were proposed and
investigated by Monte Carlo simulations in \cite{Catterall:2007fp,
  Catterall:2008yz, Catterall:2009xn}, and it was also shown that the
(naive) Wilson discretization does not require any fine tuning to
reach the correct continuum limit.  A different non-lattice approach
has been followed by \cite{Hanada:2007ti, Anagnostopoulos:2007fw,
  Hanada:2011fq, Honda:2013nfa} which used a momentum cutoff
regularization while completely fixing the gauge.

Our motivation to study the loop formulation of this model is
threefold. Apart from the motivation given by the interesting physics
related to the thermodynamics of black holes and the possibility to
test the gauge/gravity duality outlined above, the loop formulation
provides a new approach to simulate fermions on the lattice
\cite{Baumgartner:2011cm}. In contrast to standard approaches the
fermion loop formulation allows for local fermion algorithms
\cite{Wenger:2008tq}, i.e., local updates of the fermionic degrees of
freedom. The simulation algorithm applicable to the loop formulation
works for massless fermions and appears not to suffer from critical
slowing down \cite{Wenger:2008tq,Wenger:2009mi}. This is of particular
importance in the context of supersymmetric field theories with
spontaneously broken supersymmetry, since in such cases one has to
deal with a massless fermionic mode, the Goldstino fermion. The third
motivation finally stems from the fact that the fermion loop
formulation offers the potential to control the fermion sign
problem. Again, this is of particular significance in theories with
spontaneously broken supersymmery where the partition function for
periodic boundary conditions, and hence the fermion determinant (or
Pfaffian), averages to zero, since it represents the vanishing Witten
index \cite{Baumgartner:2011cm, Baumgartner:2011jw,
  Baumgartner:2012np, Baumgartner:2013ara}. The possibility to control
the fermion sign then follows from the fact that in the loop
formulation the fermionic contribution to the partition function
decomposes into contributions from fixed fermion number sectors, each
of which has a definite sign depending only on the specific choice of
the fermionic boundary conditions.

The paper is organised as follows. In section \ref{sec:lattice
  regularisation} we discuss the $d = 4$ dimensional SYM QM in the
continuum and describe the lattice regularisation using the Wilson
derivative. In section \ref{sec:fermion matrix reduction} we derive a
reduction formula for the determinant of the fermion matrix which
separates the dependence of the bosonic degrees of freedom from the
chemical potential and which then allows the straightforward
discussion of the canonical sectors of the theory. In section
\ref{sec:fermion loop formulation} the fermion loop formulation is
introduced and in section \ref{sec:fermion sectors} we discuss the
various fermion sectors emerging from the transfer matrices in the
loop formulation. We close the main part of the paper with our
conclusions and an outlook in section \ref{sec:outlook}. Finally, in
appendix \ref{app:coefficients characteristic polynomial} we review
various ways how to determine the canonical determinants from the
reduced fermion matrix and prove in appendix \ref{app:equivalence
  canonical determinants} the algebraic equivalence between the
reduced fermion matrix approach and the fermion loop formulation.

\section{Lattice regularisation}
\label{sec:lattice regularisation}
We start from $\mathcal{N}=1$ SYM in $d=4$ dimensions with gauge group
SU($N$) and dimensionally reduce the theory by compactifying the three
spatial dimensions. While the temporal component $A(t)$ of the
4-dimensional gauge field remains unchanged, the three spatial
components become bosonic fields $X_i(t), \, i=1,2,3$.  The action of
the dimensionally reduced theory then reads
\begin{equation}
  S=\frac{1}{g^2}\int_{0}^{\beta} dt \, \text{Tr} \left\lbrace  \left(  D_t X_i   \right)^2 -\frac{1}{2} \left[  X_i , X_j  \right]^2 
    +\overline{\psi} D_t \psi -\overline{\psi} \sigma_i \left[ X_i , \psi \right]
  \right\rbrace 
\end{equation}
where the anticommuting fermion fields $\overline{\psi}(t),\, \psi(t)$
are complex 2-component spinors, $\sigma_i$ are the three Pauli
matrices and $D_t=\partial_t-i[A(t),\, \cdot\,]$ denotes the covariant
derivative. All fields in the theory are in the adjoint representation
of SU($N$) and the theory possesses a ${\cal N}=4$ supersymmetry.

Note that the analogue reduction from $\mathcal{N}=1$ SYM in $d=10$
dimensions yields a very similar action with the only change that
there are 9 bosonic fields $X_i(t), \, i=1,\ldots,9$ corresponding to
the 9 compactified gauge degrees of freedom, the $\sigma_i$'s are the
SO(9) $\gamma$-matrices and the fermionic Grassmann variables are
Majorana, i.e., can be taken to be real. The dimensionally reduced
theory then corresponds to ${\cal N}=16$ SYM QM.

Let us now describe the lattice regularised version of the ${\cal
  N}=4$ SYM QM where the Euclidean time extent is discretised by $L_t$
points. The bosonic part of the action is then given by
\begin{equation}
 {S}_B = \frac{1}{g^2}\sum_{t=0}^{L_t-1} \, \Tr \left\lbrace  {\hat
     D}_t X_i(t) {\hat D}_t X_i(t)   -\frac{1}{2} \left[  X_i(t) , X_j(t)  \right]^2 \right\rbrace
\end{equation}
where the gauge field is replaced by the gauge link $U(t)$ living in
the gauge group SU($N$) and the covariant lattice derivative is
explicitly given by ${\hat D}_t X_i(t)= U(t) X_i(t+1)
U^\dagger(t)-X_i(t)$. For the regularisation of the fermionic part we
use the Wilson discretisation to get rid of the fermion doublers. Note
that in $d=1$ dimensions adding a Wilson term with Wilson parameter $r
= \pm 1$ to the symmetric derivative yields either a forward or
backward derivative,
\begin{equation}
 \partial^{\mathcal{W}} = \frac{1}{2} (\nabla^+ + \nabla^-) \pm
   \frac{1}{2} \nabla^+ \nabla^- 
= \nabla^\pm  \, .
\end{equation}
Hence, the discretised fermion action reads 
\begin{equation}
{S}_F=\frac{1}{g^2}\sum_{t=0}^{L_t-1} \, \Tr \left\lbrace
  \overline{\psi}(t) {\hat D}_t \psi(t) -\overline{\psi}(t) \sigma_i \left[ X_i(t) , \psi(t) \right] \right\rbrace 
\end{equation}
where ${\hat D}_t$ is simply the covariant derviative defined
above. Note that the Wilson term breaks the time reversal and hence
also the charge conjugation symmetry. However, the symmetries are
restored in the continuum limit together with the full supersymmetries
without any fine tuning since any further symmetry breaking terms are
prohibited by the gauge symmetry \cite{Catterall:2007fp}.

For our further discussion of the fermionic part of the theory, it is
convenient to work in uniform gauge $U(t) = U$, although it is not
necessary for the derivation of the reduced fermion matrix in the next
section. In addition, we also include a finite chemical potential term
$e^\mu$ in the forward fermion derivative \cite{Hasenfratz:1983ba} in
order to facilitate our discussion of the canonical fermion sectors in
the next section.  To be specific, the fermion action then reads
\begin{equation}
{S}_F=\frac{1}{2g^2} \sum_{t=0}^{L_t-1} \,
\left[-\overline{\psi}_\alpha^a(t) W^{ab}_{\alpha\beta} \, e^{\mu} \, \psi_\beta^b(t+1) +\overline{\psi}_\alpha^a(t)\Phi_{\alpha \beta}^{ab}(t)\psi_\beta^b(t) 
\label{eq:fermion lattice action}
 \right] 
\end{equation}
where the gauge part of the hopping term connecting the nearest
neighbour Grassmann fields $\overline{\psi}_\alpha^a(t)$ and
$\psi_\beta^b(t+1)$ is given by
\begin{equation}
\label{eq:definition W}
W^{ab}_{\alpha\beta} = 2 \delta_{\alpha\beta} \cdot\Tr\{T^a U T^b
U^\dagger\}
\end{equation}
and is independent of $t$. Here, $T^a$ are the generators of the
SU($N$) algebra and are normalised such that $\det W = 1$. The Yukawa
interaction between the fermionic and bosonic fields is described by a
$2 (N^2-1) \times 2 (N^2-1)$ matrix
\begin{equation}
\Phi_{\alpha \beta}^{ab}(t) = (\sigma_0)_{\alpha\beta} \cdot \delta^{ab} -
 2 \, (\sigma_i)_{\alpha\beta} \cdot \Tr\{T^a [X_i(t), T^b]\} 
\end{equation}
and the fermion action can be compactly written in terms of the
fermion Dirac matrix ${\cal D}_{p,a}$, i.e.,
\begin{equation}
S_F 
= \frac{1}{2g^2} \overline{\psi} \, {\cal D}_{p,a}[U, X_i;\mu] \, \psi
\, .
\end{equation}
where the subscripts $_{p,a}$ specify periodic or antiperiodic
temporal boundary conditions for the fermions in time, $\psi(L_t) =
\pm \psi(0)$, respectively.

Eventually, the grand canonical partition function reads
\begin{equation}
Z = \int{\cal D}U \, {\cal D}X_i \,  e^{-S_B[U,X_i]} \det {\cal
   D}_{p,a}[U,X_i;\mu] \, 
\end{equation}
where the determinant of the fermion Dirac matrix is the result from
integrating out the fermionic degrees of freedom $\overline{\psi}$ and
$\psi$.

\section{Fermion matrix reduction and canonical formulation}
\label{sec:fermion matrix reduction}
In $d=1$ dimensions the fermion matrix is particularly simple and
takes a cyclic block bidiagonal form,
\begin{equation}
{\cal D}_{p,a} = 
\left(
\begin{array}{ccccc}
\Phi(0) & -W  e^{\mu}     &   &   &\\
        & \Phi(1) & -W e^{\mu} &  &\\
        &        &  \Phi(2 )&  \ddots &\\
        &        &        &\ddots & -W e^{\mu} \\
 \mp W e^{\mu}  &        &        & & \Phi(L_t-1) 
\end{array}
\right) \, .
\end{equation}
Subsequently, determinant reduction techniques based on Schur
complements similar to the ones described in \cite{Alexandru:2010yb}
can be applied. As a consequence the grand canonical determinant for
the reduced fermion matrix yields
\begin{equation}
\det {\cal D}_{p,a}[U,X_i; \mu] = \det \left[ {\cal T} \mp e^{+ \mu L_t} \right]
\label{eq:reduced determinant}
\end{equation}
where ${\cal T}$ is the simple matrix product
\begin{equation}
{\cal T} = \prod_{t=0}^{L_t-1} (\Phi(t)  W) \, .
\label{eq:Tmatrix}
\end{equation}
For given background fields $U$ and $X_i(t)$ the formula allows to
calculate the determinant for any value of the chemical potential
$\mu$ by simply diagonalising ${\cal T}$ and evaluating the
characteristic polynomial of order $2(N^2-1)$ in $e^{\mu L_t}$. The
coefficients of the polynomial are then just the fermion contributions
to the grand canonical partition functions \cite{Alexandru:2010yb},
\begin{equation}
\det {\cal D}_{p,a}[U,X_i; \mu] = \sum_{n_f=0}^{2(N^2-1)} 
(\mp e^{\mu L_t})^{n_f} 
\det{\cal D}_{n_f}[U,X_i] \, ,
\label{eq:grand canonical determinant expansion}
\end{equation}
which is the conventional fugacity expansion.  Note that the
computational effort to evaluate eq.(\ref{eq:reduced determinant})
grows only linearly with the temporal extent of the lattice (through
the number of multiplications in the product), for example as one
takes the continuum limit $L_t \rightarrow \infty$. One can also work
in temporal gauge in which all gauge links are transformed to unity
except one denoted by $\widetilde W$, e.g., the one connecting time
slice $t=L_t-1$ and $t=0$. The relation to the uniform gauge is then
$\widetilde W = W^{L_t}$ and the product becomes $\prod_{t=0}^{L_t-1}
\Phi(t) \cdot \widetilde W$. Finally we note that for ordinary
supersymmetric quantum mechanics the expression for ${\cal T}$
reduces to the result given in \cite{Bergner:2007pu}.

Next we turn to the explicit evaluation of the canonical determinants.
Denoting the eigenvalues of ${\cal T}$ in eq.(\ref{eq:Tmatrix}) by
$\tau_j, j=1,\ldots,2(N^2-1)$ we can express the determinants directly
in terms of these by comparing the coefficients of the characteristic
polynomial
\begin{equation}
\det {\cal D}_{p,a}[U,X_i; \mu] = \prod_{j=1}^{2(N^2-1)} \left(\tau_j \mp  e^{\mu L_t}\right)
\end{equation}
with eq.(\ref{eq:grand canonical determinant expansion}).  The
canonical determinant in the sector with $n_f=2(N^2-1) \equiv
n_f^\text{max}$ fermions is trivial,
\begin{equation}
\det{\cal D}_{n_f^\text{max}}[U, X_i] = 1 \, ,
\label{eq:canonical determinant in the nfmax sector}
\end{equation}
which simply reflects the fact that the sector with maximally
saturated fermion number is quenched.  For the sector with $n_f = 0$
we obtain
\begin{equation}
\det{\cal D}_{n_f=0}[U, X_i] = \prod_{j=1}^{2(N^2-1)} \tau_j  = \det \left[ \prod_{t=0}^{L_t-1} (\Phi(t)
  W) \right] = \det \left[ \prod_{t=0}^{L_t-1} \Phi(t) \right]
\label{eq:canonical determinant in nf0 sector}
\end{equation}
where we made use of the fact that $\det W = 1$. The formula shows
that the fermion contribution in the $n_f=0$ sector is nontrivial,
even though it is independent of the gauge link $U$.

The sectors with $n_f=1$ and $n_f= n_f^\text{max}- 1$ fermions are
similarly simple,
\begin{eqnarray}
\label{eq:canonical determinant in nf1 sector}
\det{\cal D}_{n_f=1} &=& \sum_{j=1}^{2(N^2-1)} \prod_{k \neq j} \tau_k \, , \\
\label{eq:canonical determinant in nfmax sector}
\det{\cal D}_{n_f = n_f^\text{max} - 1} &=& \sum_{j=1}^{2(N^2-1)}
\tau_j \,\,  = \,\, \Tr({\cal T})\, .
\end{eqnarray}
The generic formula for the canonical determinants in terms of the
eigenvalues can be expressed by the elementary symmetric functions
$S_k$ of the $n_f^\text{max}$ eigenvalues
$\tau_1,\ldots,\tau_{n_f^\text{max}}$ with $k \leq
n_f^\text{max}$. The $k^{th}$ elementary symmetric function is defined
as
\begin{equation}
S_k({\cal T}) \equiv S_k(\tau_1,\ldots,\tau_{n_f^\text{max}}) = \sum_{1\leq
  i_1 < \cdots < i_k \leq n_f^\text{max}} \prod_{j=1}^k
\tau_{i_j} \, ,
\end{equation}
where the sum has $\left( {n_f^\text{max}} \atop k \right)$ summands,
and the canonical determinant in the sector with $n_f$ fermions
eventually reads
\begin{equation}
\label{eq:Dk symmetric}
\det {\cal D}_{n_f} = S_{n_f^\text{max} - n_f}({\cal T}) \, .
\end{equation}

Of course the coefficients of the characteristic polynomial can be
obtained in many other ways. In appendix \ref{app:coefficients
  characteristic polynomial} we present several alternative methods
how to calculate the canonical determinants directly from the matrix
${\cal T}$. One method makes use of the traces of powers of ${\cal T}$
while the other employs the minors of ${\cal T}$. The latter turns out
to be closely related to the transfer matrices emerging from the
fermion loop formulation discussed in the next section.

\section{Fermion loop formulation}
\label{sec:fermion loop formulation}
In the fermion loop formulation the decomposition into the various
fermion sectors are recovered in a completely different and
independent way.  The formulation is based on the exact hopping
expansion of the fermion Boltzmann factor involving the action in
eq.(\ref{eq:fermion lattice action}). Since the overall prefactor
$1/2g^2$ only contributes a trivial factor we suppress it in the
following. We apply the expansion not only to the hopping term, but in
fact to all terms in the fermion action including the Yukawa term. The
expansion is exact because it naturally truncates after the first two
terms due to the nilpotency of the Grassmann variables. Such an
expansion is most conveniently expressed by
\begin{equation}
e^x = 1 + x = \sum_{m=0}^1 x^m \, ,
\label{eq:expansion template}
\end{equation}
i.e., in terms of occupation numbers $m$.
Applying this equation
to each term in the fermion action eq.(\ref{eq:fermion lattice
  action}) characterised by the colour indices $a,b$, the Dirac
algebra indices $\alpha, \beta$ and the time coordinate $t$, the
expansion of the fermion Boltzmann factor yields
\begin{multline}
\exp(-{S}_F) = 
\prod_{t,a,b,\alpha,\beta}\left[ \sum_{m_{\alpha\beta}^{ab}(t)=0}^{1}
\left(- \Phi_{\alpha \beta}^{ab}(t)\overline{\psi}_\alpha^a(t) \psi_\beta^b(t)\right)
^{m_{\alpha \beta}^{ab}(t)} \right] \\
 \times  \prod_{t,a,\alpha}\left[ \sum_{h^{ab}_{\alpha\beta}(t)=0}^{1}
   \left(\overline{\psi}^{a}_\alpha(t) W^{ab}_{\alpha\beta} \psi^{b}_\beta(t+1) \right)
^{h^{ab}_{\alpha\beta} (t)} \right]   \, ,
\label{eq:hopping expansion}
\end{multline}
Here, the terms in the first product follow from the Yukawa
interaction while the terms in the second product stem from the
hopping terms in which we have put $\mu=0$ to simplify the
discussion. Note that one has a separate expansion for every
combination of indices $t, a, b, \alpha, \beta$ which stops after the
first two terms due to the Grassmannian character of the fermionic
degrees of freedom. The two terms in each expansion are characterised
by the occupation numbers $h^{ab}_{\alpha\beta} (t)$ and $m_{\alpha
  \beta}^{ab}(t)$ taking the values 0 or 1. The Grassmann integration
over the fermion fields requires that every pair
$\overline{\psi}^a_\alpha(t)\psi^a_\alpha(t)$ needs to be saturated by
the integration measure in order to give a nonvanishing
contribution. This condition yields local constraints on the
occupation numbers $h^{ab}_{\alpha\beta} (t)$ and $m_{\alpha
  \beta}^{ab}(t)$ separately at each site $t$,
\begin{eqnarray}
\label{eq:occupation number constraint 1}
\sum_{\alpha,a}
\left(h_{\alpha\beta}^{ab}(t-1)+m_{\alpha\beta}^{ab}(t)\right) &=& 1
    \quad \forall \beta, b, t \, , \\ 
\sum_{\beta,b} \left(h_{\alpha\beta}^{ab}(t)+m_{\alpha\beta}^{ab}(t)\right) &=& 1
    \quad \forall \alpha, a, t \, .
\label{eq:occupation number constraint 2}
\end{eqnarray}
The integration over the fermion fields is then replaced by a
summation over all configurations of occupation numbers satisfying the
constraints above.
 
\begin{figure}[h] 
\begin{center}
\begin{tikzpicture}[scale=0.8]
\node (A1) at (0,0) {};
\node (A2) at (0.5,0) {};
\node (A3) at (1,0) {};
\node (A4) at (1.5,0) {};
\node (A5) at (2,0) {};
\node (A6) at (2.5,0) {};
\node (A7) at (3,0) {};

\shade[ball color=red,opacity=.1] (A1) circle (0.1);
\shade[ball color=red,opacity=.3] (A2) circle (0.1);
\shade[ball color=red,opacity=.6] (A3) circle (0.1);
\shade[ball color=red]	 (A4) circle (0.1);
\shade[ball color=red,opacity=.6] (A5) circle (0.1);
\shade[ball color=red,opacity=.3] (A6) circle (0.1);
\shade[ball color=red,opacity=.1] (A7) circle (0.1);

\draw[->, thick]  (A4) to[out=90, in=90] (A6);
\node at (1.5,-0.5) {\tiny $\overbrace{a,\alpha}_{ }$};
\node at (2.5,-0.5) {\tiny $\overbrace{b,\beta}_{ }$};
\node at (0,-1.25)[right] {\small weight: $\Phi_{\alpha\beta}^{ab}(t)$};

\node (B1) at (5,0) {};
\node (B2) at (5.5,0) {};
\node (B3) at (6,0) {};
\node (B4) at (6.5,0) {};
\node (B5) at (7,0) {};
\node (B6) at (7.5,0) {};
\node (B7) at (8,0) {};

\shade[ball color=red,opacity=.1] (B1) circle (0.1);
\shade[ball color=red,opacity=.3] (B2) circle (0.1);
\shade[ball color=red,opacity=.6] (B3) circle (0.1);
\shade[ball color=red]	 (B4) circle (0.1);
\shade[ball color=red,opacity=.6] (B5) circle (0.1);
\shade[ball color=red,opacity=.3] (B6) circle (0.1);
\shade[ball color=red,opacity=.1] (B7) circle (0.1);

\node at (-1,0) {\small $t$};
\node at (6.5,-0.5) {\tiny $\overbrace{a,\alpha}_{ }$};

\path[->] (B4) edge[loop above,looseness=8,in=55,out=125, thick] node {} (B4);
\node at (5,-1.25)[right] {\small weight: $\Phi_{\alpha\alpha}^{aa}(t)$};
\end{tikzpicture}

\caption{\label{fig:spatial hop} Graphical representation of the
  Yukawa interaction between the fermionic degree of freedom
  characterised by $(a,\alpha)$ on time slice $t$ with the one
  characterised by $(b,\beta)$ on the same time slice and $(a,\alpha)$
  with itself (monomer term). The contributions of the interactions
  (weights) after the Grassmann integrations are also given.  }
\end{center}
\end{figure}
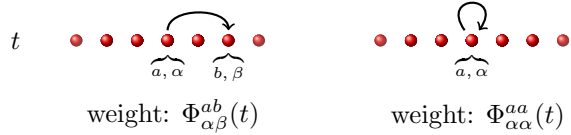

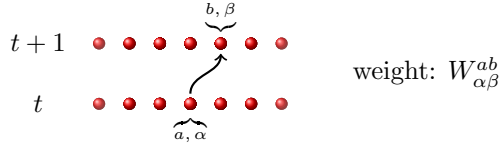
\begin{figure}[h]
\begin{center}
\begin{tikzpicture}[scale=0.8]

\node (A1) at (0,0) {};
\node (A2) at (0.5,0) {};
\node (A3) at (1,0) {};
\node (A4) at (1.5,0) {};
\node (A5) at (2,0) {};
\node (A6) at (2.5,0) {};
\node (A7) at (3,0) {};

\node (B1) at (0,1) {};
\node (B2) at (0.5,1) {};
\node (B3) at (1,1) {};
\node (B4) at (1.5,1) {};
\node (B5) at (2,1) {};
\node (B6) at (2.5,1) {};
\node (B7) at (3,1) {};

\node (To) at (1.5,1) {};
\node at (-1,0) {\small $t$};
\node at (-1,1) {\small $t+1$};

\shade[ball color=red,opacity=.1] (A1) circle (0.1);
\shade[ball color=red,opacity=.3] (A2) circle (0.1);
\shade[ball color=red,opacity=.6] (A3) circle (0.1);
\shade[ball color=red]	 (A4) circle (0.1);
\shade[ball color=red,opacity=.6] (A5) circle (0.1);
\shade[ball color=red,opacity=.3] (A6) circle (0.1);
\shade[ball color=red,opacity=.1] (A7) circle (0.1);

\shade[ball color=red,opacity=.1] (B1) circle (0.1);
\shade[ball color=red,opacity=.3] (B2) circle (0.1);
\shade[ball color=red,opacity=.6] (B3) circle (0.1);
\shade[ball color=red]	 (B4) circle (0.1);
\shade[ball color=red,opacity=.6] (B5) circle (0.1);
\shade[ball color=red,opacity=.3] (B6) circle (0.1);
\shade[ball color=red,opacity=.1] (B7) circle (0.1);

\draw[->, thick]  (A4) to[out=90, in=-90] (B5);
\node at (2,1.3) {\tiny $\underbrace{b,\beta}_{ }$};
\node at (1.5,-0.5) {\tiny $\overbrace{a,\alpha}_{ }$};

\node at (4,0.5)[right] {\small weight: $W_{\alpha\beta}^{ab}$};
\end{tikzpicture}
\caption{\label{fig:temporal hop} Graphical representation of a gauged
  temporal hop connecting the fermionic degree of freedom
  characterised by $(a,\alpha)$ on time slice $t$ with the one
  characterised by $(b,\beta)$ on time slice $t+1$. The contribution
  of the hop (weight) after the Grassmann integrations is also given.}
\end{center}
\end{figure}

The various configurations of occupation numbers and the corresponding
constraints can most easily be specified graphically by representing
each pair $\overline{\psi}^a_\alpha(t)\psi^a_\alpha(t)$ by a point
\begin{tikzpicture}
   \shade[ball color=red](0,0) circle (0.1);
\end{tikzpicture} 
and each occupation number $h^{ab}_{\alpha\beta} (t),
m_{\alpha\beta}^{ab}(t)$ by an arrow $\longrightarrow$ pointing from
point $(a, \alpha)$ to $(b, \beta)$ saturating
$\overline{\psi}^a_\alpha$ and $\psi^b_\beta$, respectively. The
graphical building blocks are then simply given by the spatial
(flavour or colour) hops characterised by $m_{\alpha\beta}^{ab}(t)=1$,
cf.~figure \ref{fig:spatial hop},

\noindent and the temporal hops characterised by $h^{ab}_{\alpha\beta}
(t)=1$, cf.~figure \ref{fig:temporal hop}, 
where the gauge links are reponsible for changing the flavour or
colour index from $a$ to $b$. Due to the breaking of the time
inversion symmetry, or equivalently charge conjugation, by the Wilson
term there exist only temporal hops in forward direction of time. The
contribution of each local fermion integration can be read off from
eq.(\ref{eq:hopping expansion}) and are given as the weights in
figures \ref{fig:spatial hop} and \ref{fig:temporal hop}. From the
contraints in eq.(\ref{eq:occupation number constraint 1}) and
(\ref{eq:occupation number constraint 2}) it becomes immediately clear
that in the graphical representation only closed, oriented fermion
loops are allowed. Moreover, each fermion loop picks up the usual
factor $(-1)$ from the Grassmann integration.  Eventually, the full
partition function in the fermion loop formulation reads
\begin{equation}
Z = \int{\cal D}U \, {\cal D}X \, e^{-S_B[U,X_i]} 
\sum_{\{h,m\}}
\prod_t \left[
  \left(W_{\alpha\beta}^{ab}\right)^{h_{\alpha\beta}^{ab}(t)}
  \left(\Phi_{\alpha\beta}^{ab}(t)\right)^{m_{\alpha\beta}^{ab}(t)} \right]
\end{equation}
where the sum is over all combinations of occupation numbers
satisfying eq.(\ref{eq:occupation number constraint 1}) and
(\ref{eq:occupation number constraint 2}).

\section{Fermion sectors and transfer matrices}
\label{sec:fermion sectors}
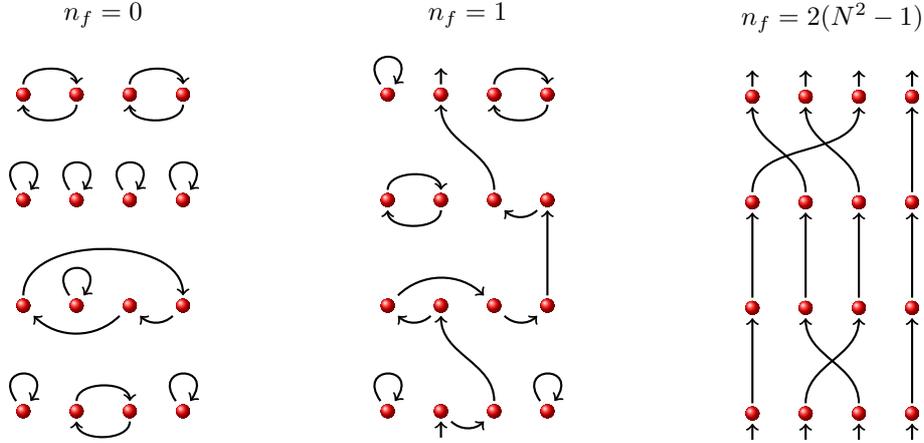
\begin{figure}[ht]
\begin{center}
\begin{minipage}[t]{0.3\textwidth}
\centering
\begin{tikzpicture}[scale=0.7]

\node (A1) at (0,0) {};
\node (A2) at (1,0) {};
\node (A3) at (2,0) {};
\node (A4) at (3,0) {};

\node (B1) at (0,2) {};
\node (B2) at (1,2) {};
\node (B3) at (2,2) {};
\node (B4) at (3,2) {};

\node (C1) at (0,4) {};
\node (C2) at (1,4) {};
\node (C3) at (2,4) {};
\node (C4) at (3,4) {};

\node (D1) at (0,6) {};
\node (D2) at (1,6) {};
\node (D3) at (2,6) {};
\node (D4) at (3,6) {};

\node at (1.5,7.5) {\small $n_f = 0$};

\shade[ball color=red] (A1) circle (0.13);
\shade[ball color=red] (A2) circle (0.13);
\shade[ball color=red] (A3) circle (0.13);
\shade[ball color=red] (A4) circle (0.13);

\shade[ball color=red] (B1) circle (0.13);
\shade[ball color=red] (B2) circle (0.13);
\shade[ball color=red] (B3) circle (0.13);
\shade[ball color=red] (B4) circle (0.13);

\shade[ball color=red] (C1) circle (0.13);
\shade[ball color=red] (C2) circle (0.13);
\shade[ball color=red] (C3) circle (0.13);
\shade[ball color=red] (C4) circle (0.13);

\shade[ball color=red] (D1) circle (0.13);
\shade[ball color=red] (D2) circle (0.13);
\shade[ball color=red] (D3) circle (0.13);
\shade[ball color=red] (D4) circle (0.13);

\path[->] (A1) edge [loop above,looseness=8,in=55,out=125, thick] node {} (A1);
\draw[->, thick]  (A2)    to[out=90, in=90] (A3);
\draw[->, thick]  (A3)    to[out=-90, in=-90] (A2);
\path[->] (A4) edge [loop above,looseness=8,in=55,out=125, thick] node {} (A4);

\path[->] (B2) edge [loop above,looseness=8,in=55,out=125, thick] node {} (B2);
\draw[->, thick]  (B1)    to[out=90, in=90] (B4);
\draw[->, thick]  (B4)    to[out=-135, in=-45] (B3);
\draw[->, thick]  (B3)    to[out=-135, in=-45] (B1);

\path[->] (C1) edge [loop above,looseness=8,in=55,out=125, thick] node {} (C1);
\path[->] (C2) edge [loop above,looseness=8,in=55,out=125, thick] node {} (C2);
\path[->] (C3) edge [loop above,looseness=8,in=55,out=125, thick] node {} (C3);
\path[->] (C4) edge [loop above,looseness=8,in=55,out=125, thick] node {} (C4);

\draw[->, thick]  (D1)    to[out=90, in=90] (D2);
\draw[->, thick]  (D2)    to[out=-90, in=-90] (D1);
\draw[->, thick]  (D3)    to[out=90, in=90] (D4);
\draw[->, thick]  (D4)    to[out=-90, in=-90] (D3);

\end{tikzpicture}
\end{minipage}
\begin{minipage}[t]{0.3\textwidth}
\centering
\begin{tikzpicture}[scale=0.7]

\node (A1) at (0,0) {};
\node (A2) at (1,0) {};
\node (A3) at (2,0) {};
\node (A4) at (3,0) {};

\node (B1) at (0,2) {};
\node (B2) at (1,2) {};
\node (B3) at (2,2) {};
\node (B4) at (3,2) {};

\node (C1) at (0,4) {};
\node (C2) at (1,4) {};
\node (C3) at (2,4) {};
\node (C4) at (3,4) {};

\node (D1) at (0,6) {};
\node (D2) at (1,6) {};
\node (D3) at (2,6) {};
\node (D4) at (3,6) {};

\node at (1.5,7.5) {\small $n_f=1$};

\shade[ball color=red] (A1) circle (0.13);
\shade[ball color=red] (A2) circle (0.13);
\shade[ball color=red] (A3) circle (0.13);
\shade[ball color=red] (A4) circle (0.13);

\shade[ball color=red] (B1) circle (0.13);
\shade[ball color=red] (B2) circle (0.13);
\shade[ball color=red] (B3) circle (0.13);
\shade[ball color=red] (B4) circle (0.13);

\shade[ball color=red] (C1) circle (0.13);
\shade[ball color=red] (C2) circle (0.13);
\shade[ball color=red] (C3) circle (0.13);
\shade[ball color=red] (C4) circle (0.13);

\shade[ball color=red] (D1) circle (0.13);
\shade[ball color=red] (D2) circle (0.13);
\shade[ball color=red] (D3) circle (0.13);
\shade[ball color=red] (D4) circle (0.13);

\path[->] (A1) edge [loop above,looseness=8,in=55,out=125, thick] node {} (A1);
\draw[->, thick]  (A2)    to[out=-45, in=-135] (A3);
\draw[->, thick]  (1,-0.5)    to[out=90, in=-90] (A2);
\path[->] (A4) edge [loop above,looseness=8,in=55,out=125, thick] node {} (A4);
\draw[->, thick]  (A3)    to[out=90, in=-90] (B2);

\draw[->, thick]  (B1)    to[out=45, in=135] (B3);
\draw[->, thick]  (B2)    to[out=-135, in=-45] (B1);
\draw[->, thick]  (B3)    to[out=-45, in=-135] (B4);
\draw[->, thick]  (B4)    to[out=90, in=-90] (C4);

\draw[->, thick]  (C4)    to[out=-135, in=-45] (C3);
\draw[->, thick]  (C1)    to[out=90, in=90] (C2);
\draw[->, thick]  (C2)    to[out=-90, in=-90] (C1);
\draw[->, thick]  (C3)    to[out=90, in=-90] (D2);

\path[->] (D1) edge [loop above,looseness=8,in=55,out=125, thick] node {} (D1);
\draw[->, thick]  (D2)    to[out=90, in=-90] (1,6.5);
\draw[->, thick]  (D3)    to[out=90, in=90] (D4);
\draw[->, thick]  (D4)    to[out=-90, in=-90] (D3);

\end{tikzpicture}
\end{minipage}
\begin{minipage}[t]{0.3\textwidth}
\centering
\begin{tikzpicture}[scale=0.7]

\node (A1) at (0,0) {};
\node (A2) at (1,0) {};
\node (A3) at (2,0) {};
\node (A4) at (3,0) {};

\node (B1) at (0,2) {};
\node (B2) at (1,2) {};
\node (B3) at (2,2) {};
\node (B4) at (3,2) {};

\node (C1) at (0,4) {};
\node (C2) at (1,4) {};
\node (C3) at (2,4) {};
\node (C4) at (3,4) {};

\node (D1) at (0,6) {};
\node (D2) at (1,6) {};
\node (D3) at (2,6) {};
\node (D4) at (3,6) {};

\node at (1.5,7.5) {\small $n_f = 2(N^2-1)$};

\shade[ball color=red] (A1) circle (0.13);
\shade[ball color=red] (A2) circle (0.13);
\shade[ball color=red] (A3) circle (0.13);
\shade[ball color=red] (A4) circle (0.13);

\shade[ball color=red] (B1) circle (0.13);
\shade[ball color=red] (B2) circle (0.13);
\shade[ball color=red] (B3) circle (0.13);
\shade[ball color=red] (B4) circle (0.13);

\shade[ball color=red] (C1) circle (0.13);
\shade[ball color=red] (C2) circle (0.13);
\shade[ball color=red] (C3) circle (0.13);
\shade[ball color=red] (C4) circle (0.13);

\shade[ball color=red] (D1) circle (0.13);
\shade[ball color=red] (D2) circle (0.13);
\shade[ball color=red] (D3) circle (0.13);
\shade[ball color=red] (D4) circle (0.13);


\draw[->, thick]  (A1)    to[out=90, in=-90] (B1);
\draw[->, thick]  (A2)    to[out=90, in=-90] (B3);
\draw[->, thick]  (A3)    to[out=90, in=-90] (B2);
\draw[->, thick]  (A4)    to[out=90, in=-90] (B4);

\draw[->, thick]  (B1)    to[out=90, in=-90] (C1);
\draw[->, thick]  (B2)    to[out=90, in=-90] (C2);
\draw[->, thick]  (B3)    to[out=90, in=-90] (C3);
\draw[->, thick]  (B4)    to[out=90, in=-90] (C4);

\draw[->, thick]  (C1)    to[out=90, in=-90] (D3);
\draw[->, thick]  (C2)    to[out=90, in=-90] (D1);
\draw[->, thick]  (C3)    to[out=90, in=-90] (D2);
\draw[->, thick]  (C4)    to[out=90, in=-90] (D4);

\draw[->, thick]  (0,-0.5)    to[out=90, in=-90] (A1);
\draw[->, thick]  (1,-0.5)    to[out=90, in=-90] (A2);
\draw[->, thick]  (2,-0.5)    to[out=90, in=-90] (A3);
\draw[->, thick]  (3,-0.5)    to[out=90, in=-90] (A4);

\draw[->, thick]  (D1)    to[out=90, in=-90] (0,6.5);
\draw[->, thick]  (D2)    to[out=90, in=-90] (1,6.5);
\draw[->, thick]  (D3)    to[out=90, in=-90] (2,6.5);
\draw[->, thick]  (D4)    to[out=90, in=-90] (3,6.5);

\end{tikzpicture}
\end{minipage}
\caption{Three sample configurations of closed oriented loops for four
  fermionic degrees of freedom (representative for the generic
  $2(N^2-1)$ ones) on a periodic lattice with four time slices.}
\label{fig:loop configurations}
\end{center}
\end{figure}

In figure \ref{fig:loop configurations} we show three sample
configurations consisting of closed oriented fermion loops for four
fermionic degrees of freedom (representative for the generic
$2(N^2-1)$ ones). One immediately notices that the configurations can
be classified according to the number of fermions $n_f$ propagating
forward in time.
For the three examples depicted in figure \ref{fig:loop
  configurations} the fermion numbers are $n_f=0, 1$ and 4
(i.e.~$n_f=2(N^2-1)$ for the generic case), respectively. In each
sector, the propagation of the $n_f$ fermions can be described by
transfer matrices
\begin{equation}
T_{n_f}(t) = T_{n_f}^\Phi\left(X_i(t)\right) \cdot T_{n_f}^W(U)
\end{equation}
where the first transfer matrix describes the various ways how to
connect $n_f$ fermions entering at time $t$ with $n_f$ fermions
exiting at $t$. It depends on the boson field configuration $X_i(t)$
through the Yukawa interactions matrix $\Phi(t)$ and hence depends on
$t$. The second transfer matrix describes how to connect $n_f$
fermions exiting at $t$ and entering at $t+1$, and hence depends on
the gauge field $U$ through $W$ in eq.(\ref{eq:definition W}). In
uniform gauge, this transfer matrix has no time dependence.  Then, for
a given gauge and boson field background $\{U, X_i(t)\}$ the fermion
contribution to the partition function in the sector with $n_f$
fermions is simply given by
\begin{equation}
\det{\cal D}_{n_f}[U,X_i] = \Tr\left[ \prod_{t=0}^{L_t-1}T_{n_f}(t) \right] \, .
\end{equation}
The full contribution is then obtained by adding up all these terms
taking into account a factor $(\mp 1) e^{\mu L_t}$ for each fermion
loop winding around the lattice in temporal direction, with the sign
depending on whether periodic or antiperiodic boundary conditions are
employed. The expression eventually reads
\begin{equation}
\det {\cal D}_{p,a}[U,X_i;\mu] = \sum_{n_f=0}^{2(N^2-1)} (\mp e^{\mu L_t})^{n_f}  \Tr\left[ \prod_{t=0}^{L_t-1}T_{n_f}(t) \right] 
\end{equation}
and can directly be compared with eq.(\ref{eq:grand canonical
  determinant expansion}).

Let us now look in more detail at the transfer matrices separately in
each sector. First we note that the size of $T_{n_f}$ is given by the
number of states in sector $n_f$, i.e.,
\begin{equation}
n \equiv \left( 
\begin{array}{c}
2(N^2-1) \\
n_f
\end{array}
\right) \, .
\end{equation}
The sectors with $n_f=0$ and $n_f=2(N^2-1)$ are therefore particularly
simple since in these cases the transfer matrix is just $1\times
1$. We will hence first discuss these two sectors, followed by the
still rather simple sectors with $n_f=1$ and $n_f=2(N^2-1) - 1$,
before presenting the generic case for arbitrary values of $n_f$.

\subsection{Sector $n_f=0$}
For $n_f=0$ we see by inspection of the corresponding configuration in
figure \ref{fig:loop configurations} that there is no gauge link
dependence, and hence $T_0^W = 1$, while the transfer matrix
$T_0^\Phi(t)$ must contain the sum of the weights of all fermion loop
configurations on a given time slice $t$. By doing so, we need to take
care that each nontrivial fermion loop picks up the usual factor
$(-1)$ from the Grassmann integration.  It is not difficult to see
that a given time slice configuration can be specified by a
permutation $\sigma$ of the indices $i=1,\ldots, 2(N^2-1)$ labelling
the fermionic degrees of freedom. Each cycle $(ijk\ldots l)$ in the
permutation then corresponds to a sequence of indices characterising a
specific fermion loop and its weight is given by $\Phi_{ij} \Phi_{jk}
\ldots \Phi_{li}$. The total sign of the configuration is given by
including a factor $(-1)$ for each nontrivial cycle, i.e., counting
whether the number of nontrivial cycles in the permutation is even or
odd which corresponds to the parity of the permutation. Finally, the
sum over all configurations amounts to summing up all permutations
including the corresponding weights and the signs given by the parity
of the permutation. This prescription is of course nothing else than
the definition for the determinant, so the transfer matrix in the
$n_f=0$ sector is simply given by
\begin{equation}
T_0^\Phi(t) = \det \Phi(t)
\end{equation}
and the total fermion contribution  factorises completely,
\begin{equation}
\det{\cal D}_{n_f=0}[U, X_i] = \prod_{t=0}^{L_t-1} \det \Phi(t) \, .
\end{equation}
Comparing this with eq.(\ref{eq:canonical determinant in nf0 sector})
we obviously find complete agreement. In the fermion loop approach
however it is evident from the beginning that the gauge link $U$ does
not contribute in the $n_f=0$ sector.

\subsection{Sector $n_f=n_f^\text{max}$}
For $n_f=2(N^2-1)\equiv n_f^\text{max}$ the transfer matrix
$T_{n_f^\text{max}}(t)$ is again $1 \times 1$.  While there are no
contributions from the Yukawa interaction, hence
$T_{n_f^\text{max}}^\Phi(t)=1$, we need to take into account the
nontrivial hopping in colour space. The complication arising here
stems from the fact that depending on the number of hoppings in colour
space, the total number of fermion loops winding in temporal direction
changes, but not the number of winding fermions. For example, if there
are only colour diagonal hops, the number of winding loops is
$n_f^\text{max}$ and the corresponding contribution comes with a
positive sign. On the other hand, if there is one single nondiagonal
colour hop two loops merge into one, so the number of winding loops
becomes $n_f^\text{max}-1$ and the contribution should hence contain a
negative sign relative to the contribution with $n_f^\text{max}$
loops. So for every nondiagonal colour hop the number of loops is
changing by one.

Similarly to the $n_f=0$ sector we need to take all permutations of
the colour indices $a,b$ into account. For each nontrivial permutation
of two indices the number of fermion loops winding in temporal
direction is reduced by one and we take this into account by including
a factor $(-1)$. Summing over all permutations including the sign
corresponding to the parity of the permutations again yields the
determinant, i.e.,
\begin{equation}
T_{n_f^\text{max}}^W = \det \left[W\right] = 1
\end{equation}
yielding the total contribution
\begin{equation}
\det{\cal D}_{n_f^\text{max}}[U, X_i] = \prod_{t=0}^{L_t-1} T_{n_f^\text{max}}(t) =  1 \, .
\end{equation}
This is in accordance with the result from the determinant reduction,
cf.~eq.(\ref{eq:canonical determinant in the nfmax sector}), and it is
obvious that the same result would be obtained without referring to a
particular gauge. Since the fermions are completely saturated by the
temporal hopping terms and contribute only trivially to the canonical
determinant, this sector corresponds to the quenched one as noted
before.

\subsection{Sector $n_f=1$}
Next, we look at the sector with $n_f=1$ fermions. The corresponding
transfer matrices $T_1(t)$ are of size $2(N^2-1) \times
2(N^2-1)$. Each matrix element $(T_1^\Phi(t))_{ij}$ contains the sum
of weights of all configurations at fixed $t$ where the fermion degree
of freedom $i=(a,\alpha)$ is entering time slice $t$ and $j=(b,\beta)$
is leaving. The corresponding degrees of freedom are then already
saturated by the corresponding hops in and out of the time slice and
hence the weights $\Phi_{ki}$ and $\Phi_{jk}, k=1,\ldots,2(N^2-1)$ can
not appear in any of the configurations. The remaining time slice
configurations can be obtained in analogy to the considerations in the
$n_f=0$ sector, that is by constructing all permutations, i.e.,~cycles
of the remaining degrees of freedom and taking into account factors of
$(-1)$ for each nontrivial cycle. Following the arguments from the
$n_f=0$ sector it turns out that this is again equivalent to taking
the determinant of $\Phi(t)$, but with row $j$ and column $i$ removed,
i.e.,
\begin{equation}
\left(T_1^\Phi\right)_{ij} = (-1)^{i+j} \left. \det \Phi
  \right|_{\Phi_{ki}=\delta_{kj},\Phi_{jk}=\delta_{ik}}
\equiv  (-1)^{i+j} \det \Phi^{ \backslash
  \hspace{-0.1cm}j \backslash \hspace{-0.1cm}i} \, 
\end{equation}
which is in fact the $(j,i)$-cofactor of $\Phi$. This will be
discussed in more detail in section \ref{sec:generic sector}.
Similarly, in order to include the colour changing hops due to the
gauge link between time slices we multiply with the corresponding
gauge link transfer matrix $T_1^W$
\begin{equation}
\left(T_1^W\right)_{ij} =  (W)_{ij} \, 
\end{equation}
which in uniform gauge is constant in
time and is in fact the complementary $(i,j)$-minor $\det W^{ij}$.
Eventually, the full fermion contribution in the $n_f=1$ sector reads
\begin{equation}
\det{\cal D}_{n_f=1}[U, X_i] = \Tr \prod_{t=0}^{L_t-1}\left[T_1^\Phi(t) \cdot T_1^W \right] \, 
\end{equation}
and comparing this result to the one in eq.(\ref{eq:canonical
  determinant in nf1 sector}) from the fugacity expansion, we find a
nontrivial relationship between the two expressions.  We will comment
further on this relation in section \ref{sec:generic sector} and
establish it in detail in appendix \ref{app:equivalence canonical
  determinants}.

\subsection{Sector $n_f=n_f^\text{max}-1$}
In the sector where all but one, i.e., $n_f^\text{max}-1$ fermions are
propagating, the states of the transfer matrices
$T_{n_f^\text{max}-1}(t)$ are most conveniently labelled by the degree
of freedom $i=(a,\alpha)$ not occupied by a temporal hopping term. The
transfer matrices are hence of size $2(N^2-1) \times 2(N^2-1) =
n_f^\text{max} \times n_f^\text{max}$. The matrix elements
$(T_{n_f^\text{max}-1}^\Phi)_{ij}$ are calculated following the
arguments outlined above for the $n_f=0$ and 1 sector, namely to take
the determinant of the Yukawa matrix $\Phi$ with all columns and rows
deleted except $i$ and $j$, respectively. The reduced Yukawa matrix is
then just a single element and hence we have
\begin{equation}
(T_{n_f^\text{max}-1}^\Phi)_{ij} = (-1)^{i+j} \Phi_{ij} \, 
\end{equation}
which is just the complementary $(i,j)$-cofactor of $\Phi$ up to an
overall sign.  The transfer matrix describing all the possible
configurations within a time slice needs to be complemented by the one
inducing the colour changing hops due to the gauge link between the
time slices. If fermion $i$ is not hopping out of $t$ and $j$ not into
$t+1$ they will not contribute, while the mixing of the remaining
degrees of freedom is described as before by taking the determinant of
the hop matrix,
\begin{equation}
(T^W_{n_f^\text{max}-1})_{ij} = \det W^{\backslash
  \hspace{-0.1cm}i \backslash \hspace{-0.1cm}j} \, 
\end{equation}
which is the $(i,j)$-minor of $W$.  The full fermion contribution in
the $n_f = n_f^\text{max}-1$ sector finally yields
\begin{equation}
\det{\cal D}_{n_f^\text{max}-1}[U,X_i] = \Tr \prod_{t=0}^{L_t-1}\left[T_{n_f^\text{max}-1}^\Phi(t) \cdot T_{n_f^\text{max}-1}^W \right] \, .
\end{equation}
This can be compared to the one in eq.(\ref{eq:canonical determinant
  in nfmax sector}) from the fugacity expansion and we find again a
nontrivial relationship between the two expressions.

\subsection{Sector with generic $n_f$}
\label{sec:generic sector}
Similar constructions can be worked out in all the other sectors, but
the constructions become more involved since the number of states
grows rapidly towards the half-filled sector with $n_f =
2(N^2-1)/2$. However, our previous discussion indicates a generic
pattern which becomes clear after careful further investigation of all
the weights and signs of each configuration. Employing some higher
linear algebra one can eventually formulate the following rule. The
sector with $n_f$ fermions contains $n=\left( {n_f^\text{max}} \atop
  {n_f}\right)$ states and the elements of the corresponding $n\times
n$ transfer matrix $T_{n_f}^\Phi$ are given by the cofactors of $\Phi$
of order $n_f$, while the matrix elements of $T^W_{n_f}$ are given by
the complementary minors of $W$.

To be more precise, let $A$ and $B$ be two index sets $A, B \subseteq
\{1,2,\ldots,2(N^2-1)\}$ of size $n_f$, then the cofactor of $\Phi$ of
order $n_f$ is the signed determinant of the $(2(N^2-1)-n_f)\times
(2(N^2-1)-n_f)$ submatrix $\Phi^{\,\backslash \hspace{-0.2cm} B
  \,\backslash \hspace{-0.2cm}A}$ obtained from $\Phi$ by deleting the
rows indexed by $B$ and the columns indexed by $A$, so
\begin{equation}
\left(T^\Phi_{n_f}\right)_{AB} = (-1)^{p(A,B)} \det  \Phi^{\, \backslash \hspace{-0.2cm} B \, \backslash
  \hspace{-0.2cm}A} \, 
\label{eq:TPhi}
\end{equation}
where $p(A,B) = \sum_{i\in A} i + \sum_{j\in B}j$, while the
complementary minor $\det W^{AB}$ is the determinant of the $n_f
\times n_f$ submatrix $W^{AB}$ obtained from $W$ by keeping only the
rows indexed by $A$ and the columns indexed by $B$,
\begin{equation}
\left(T^W_{n_f}\right)_{AB} = \det W^{AB} \, .
\label{eq:TW}
\end{equation}
If the two sets $A$ and $B$ are equal, the cofactors reduce to minors
and the corresponding determinants are called principal minors or
principal complementary minors. Note also that in the literature the
role of the minor and complementary minor is sometimes exchanged.

In analogy to the discussion before, the cofactor $C_{\,\backslash
  \hspace{-0.2cm} B \, \backslash \hspace{-0.2cm}A}(\Phi) =
(-1)^{p(A,B)} \det \Phi^{\,\backslash \hspace{-0.2cm} B \, \backslash
  \hspace{-0.2cm}A}$ includes all contributions to the transition of
$n_f$ fermions indexed by $A$ entering at time $t$ to $n_f$ fermions
indexed by $B$ exiting from time $t$, with all the weights and signs
properly accounted for. Similarly, the minor $M_{AB}(W) = \det W^{A
  B}$ connects $n_f$ fermions indexed by $A$ exiting $t$ in all
possible ways with $n_f$ fermions indexed by $B$ entering time $t+1$
with the correct weight and sign for each connection.  Hence, the full
transfer matrix at time $t$ in the sector with $n_f$ fermions is then
$T_{n_f}^\Phi(t) \cdot T^W_{n_f}$ and the corresponding canonical
determinant reads
\begin{equation}
\label{eq:generic canonical det}
\det {\cal D}_{n_f}[U,X_i] = \Tr \prod_{t=0}^{L_t-1}
\left[T_{n_f}^\Phi(t) \cdot T^W_{n_f}\right] \, .
\end{equation}
It is easy to check that this generic definition yields the correct
expressions for the transfer matrices and canonical determinants for
the cases $n_f=0,1,n_f^\text{max}-1,n_f^\text{max}$ discussed in the
previous sections. Note that for the empty sets $A=B=\{\}$ the
principal minor, and analogously the complementary principal minor for
the full sets $A=B=\{1,\ldots,2(N^2-1)\}$, is 1 by definition.

Finally, one can show that the canonical determinants obtained in the
fermion loop approach are equal to the ones using the fermion matrix
reduction, cf.~eq.(\ref{eq:Dk symmetric}). Using various relations
between matrices of minors and cofactors, one can derive that
\begin{equation}
\left(\prod_{t=0}^{L_t-1} \left[T_{n_f}^\Phi(t) \cdot T_{n_f}^W
\right]\right)_{AB} = (-1)^{p(A,B)} \det {\cal T}^{\,\backslash
  \hspace{-0.2cm} A  \, \backslash \hspace{-0.2cm}B} = C_{\,\backslash
  \hspace{-0.2cm} A  \, \backslash \hspace{-0.2cm}B}({\cal T}) \, .
\end{equation}
The details of this derivation are given in appendix
\ref{app:equivalence canonical determinants}.  The r.h.s.~forms the $n
\times n$ matrix of cofactors of ${\cal T}$ of order $n_f$ and the
trace in eq.(\ref{eq:generic canonical det}) then yields the sum over
the $n$ principal minors of ${\cal T}$ of order $n_f$ denoted by
$E_{n_f}$, i.e.,
\begin{equation}
\label{eq:Dk minors}
\det {\cal D}_{n_f} = \sum_{B} \det {\cal T}^{\,\backslash \hspace{-0.2cm} B  \, \backslash
  \hspace{-0.2cm}B}  \equiv E_{n_f}({\cal T}) \, .
\end{equation}
Recalling a known relation from linear algebra between the sum of
minors of a matrix and its symmetric functions \cite{Horn:2013} one
has
\begin{equation}
E_{n_f}({\cal T}) = S_{n_f^\text{max}-n_f}({\cal T})
\end{equation}
which establishes the equivalence between eq.(\ref{eq:Dk symmetric})
and eq.(\ref{eq:Dk minors}).

\subsection{Remarks}
We close this section with several remarks.  Firstly, we note that in
contrast to the full determinant $\det[U, X_i]$, which can be proven
to be positive \cite{Catterall:2007fp}, the various canonical
determinants $\det{\cal D}_{n_f}[U, X_i]$ need not necessarily be
positive. Obviously, $\det{\cal D}_{n_f=2(N^2-1)}[U, X_i]$ is so and
it seems that at least $\det{\cal D}_{n_f=0}[U, X_i]$ is also
positive, although we do not have any proof. It would be interesting
to study potential fermion sign problems in the canonical sectors in
the present model. Despite its simplicity due to the low dimension, it
nevertheless contains all the important features of a gauge theory,
and hence conclusions can most likely be generalised to more
complicated gauge theories in higher dimensions, such as QCD in the
canonical formulation \cite{Alexandru:2010yb}.

Secondly, we note that the various sectors, in particular the ones
with many fermions, can in principle be simulated by open fermion
string (fermion worm) algorithms along the lines described in
\cite{Wenger:2008tq,Wenger:2009mi}. This approach has indeed already
been applied successfully in ordinary supersymmetric quantum mechanics
\cite{Baumgartner:2011cm}, in the supersymmetric nonlinear O($N$)
sigma model \cite{Steinhauer:2013tba} and in the two dimensional
${\cal N}=1$ Wess-Zumino model
\cite{Baumgartner:2011jw,Baumgartner:2013ara} where the transfer
matrix techniques discussed here and in \cite{Baumgartner:2012np} are
out of reach. Furthermore, for the model discussed in this paper, a
discrete bond formulation for the bosonic degrees of freedom is
available \cite{PhD_Steinhauer:2014}. Such a formulation promises a
huge gain in efficiency for numerical simulations, but it is not clear
whether the bosonic bond formulation can be put into practice.

Thirdly, from investigations in the Hamiltonian formulation of the
theory \cite{Wosiek:2002nm}, where time is treated as a continuous
variable, it is known that there is a (spectral) symmetry between the
sectors with $n_f$ and $2(N^2-1)-n_f$ fermions, due to the exchange
symmetry between particles and antiparticles. Our results above
indicate that the symmetry is not maintained by our choice of the
discretisation in the Lagrangian formalism, but the reason for this is
clear. As we mentioned earlier the Wilson term needed to control the
doubler fermions explicitly breaks the time reversal and hence the
charge conjugation symmetry which of course is crucial for an exact
particle/antiparticle exchange symmetry. However, since the symmetries
are restored in the continuum limit without fine tuning, the
symmetries between the various canonical sectors will also be mended
automatically in the continuum, and the difference between the related
sectors will provide a good estimate of the remaining systematic
lattice artefacts.

\section{Conclusions and outlook}
\label{sec:outlook}
In this paper we have investigated in detail the structure of the
fermionic part of the $d=4$ dimensional supersymmetric Yang-Mills
quantum mechanics, i.e., ${\cal N}=4$ SYM QM with gauge group
SU($N$). On the one hand, we derived a reduced fermion matrix whose
size is independent of the temporal extent of the lattice. In
addition, the dependence on the chemical potential is factored out and
this allows the exact projection of the fermion determinant onto the
canonical sectors with fixed fermion number, once the eigenvalues of
the reduced matrix are calculated. On the other hand, we have
presented the fermion loop formulation of the theory in which the
grand canonical fermion determinant naturally decomposes into sectors
with fixed fermion numbers. The construction of transfer matrices is
rather straightforward in the various fermion sectors and the
comparison with the fugacity expansion, accessible via the reduced
fermion matrix, yields identical results and interesting relations
between the transfer matrices and the eigenvalues of the reduced
fermion matrix. In fact, we presented a proof which establishes the
equivalence of the canonical determinants from the reduced fermion
matrix approach and from the fermion loop formulation on the algebraic
level.

Our results open various possibilities for a range of nonperturbative
investigations of the theory. This can be done for example by
numerical simulations using methods different from the usual Hybrid
Monte Carlo approach, either using the transfer matrices in the
various canonical sectors with fixed fermion numbers, or using the
projection to the sectors with the help of the reduced fermion
matrix. Another interesting approach could be the application of mean
field methods to the spatial gauge degrees of freedom, again either in
the transfer matrix approach or using the reduced fermion matrix. It
is even conceivable that the methods presented here and the emerged
simple structures lead to new analytic results in some interesting
limits. All results obtained either way will provide important
insights into the conjectured M-theory and will add to our
understanding of the corresponding gauge/gravity duality, besides
unveiling interesting physics of the model itself.

Another interesting line of research starting from here concerns the
investigation of ordinary, non-supersymmetric gauge field theories in
higher dimensions at finite fermion density, such as QCD at finite
baryon density. It is notoriously difficult to obtain reliable results
in these theories using the known numerical approaches, due to the
intrinsic fermion sign problem at finite density, and any insight into
how the simulations of these theories could be facilitated would be
extremely valuable. The explicit fugacity expansion derived in this
paper allows to investigate finite density simulations or canonical
simulations in a simple setup which nevertheless displays a similar
structure, and hence contains all the important features, as the more
complicated theories in higher dimensions such as QCD.

Finally, the extension of the loop formulation to ${\cal N} = 16$
supersymmetric Yang-Mills quantum mechanics is in principle
straightforward but requires special care. This is due to the fact
that the corresponding dimensionally reduced model has obviously a
different Dirac structure, and it remains to be seen whether the
structure is compatible with the requirements for the fermion loop
formulation. The fermion matrix reduction on the other hand should be
unaffected by the change of the Dirac structure.

\section*{Acknowledgements}
We thank Dan Boss, Adolpho Guarino, Piotr Korcyl, Andreas Wipf and
Jacek Wosiek for useful discussions, and in particular Piotr Korcyl
and Dan Boss for sharing their codes for the $d=4$ dimensional SU($N$)
SYM QM with $N=2$ and $N=2,3$, respectively.

\begin{appendix}
\section{Determinations of canonical determinants}
 \label{app:coefficients characteristic polynomial}
 In this appendix we review three alternative methods to calculate the
 canonical determinants from the matrix ${\cal T}$ in
 eq.(\ref{eq:Tmatrix}). As shown in section \ref{sec:fermion matrix
   reduction} the canonical determinants are just the coefficients of
 the characteristic polynomial of the matrix ${\cal T}$. The first
 method provides recursion relations which yield the coefficients in
 terms of the eigenvalues $\tau_i$ of ${\cal T}$. The second method
 evaluates the coefficients in terms of the traces of powers of ${\cal
   T}$ and the third makes use of the minors of ${\cal T}$. The latter
 method turns out to be closely related to the transfer matrix
 approach in the fermion loop formulation and hence deserves special
 emphasis.

 In the following we assume the matrix ${\cal T}$ to be of size $n
 \times n$ and for simplicity we consider only the case of
 antiperiodic b.c., hence the relevant characteristic polynomial is
\begin{equation}
g(x) = \det({\cal T} + x \cdot \mathbbm{1}) =  \sum_{k=0}^{n} c_k \cdot x^k
\end{equation}
where $\mathbbm{1}$ is the $n \times n$ unit matrix and the
coefficients $c_k$ are the canonical determinants $\det{\cal
  D}_{n_f=k}$ in sector $k$.

\subsection{Coefficients from recursion relations}
The coefficients can be obtained from the eigenvalues $\tau_i$ of
${\cal T}$ using recursive relations \cite{Alexandru:2010yb}. To this
end, we first define the partial products
\begin{equation}
\Pi_r(x) = \prod_{j=1}^{r} (\tau_j + x) = \sum_{k=1}^{r} c_k^{(r)} x^k 
\end{equation}
which fulfill $\prod_{r+1}(x) = (\tau_{r+1} + x) \prod_r(x)$. Setting
$c_{-1}^{(r)} = 0$ we have the recursion relation
\begin{equation}
c_k^{(r+1)} = \tau_{r+1} c_k^{(r)} + c_{k-1}^{(r)}
\end{equation}
for all $0 \leq k \leq r+1$ which allows to compute $c_k^{(r+1)}$ from
$c_k^{(r)}$. After $n$ steps we obtain the coefficients $c_k \equiv
c_k^{(n)}$ of $\prod_{n}(x)$ which are then just the canonical
determinants $\det{\cal D}_{n_f = k}$. The generalisation of the
recursion to include the minus sign from the periodic b.c.~is
straightforward.

\subsection{Coefficients in terms of traces}
\label{app:coefficients from traces}
Here we review the calculation of the coefficients $c_k$ in terms of
traces of powers of the matrix ${\cal T}$. To do so we introduce the
notation
\begin{equation}
t_k = \Tr({\cal T}^k) \, .
\end{equation}
Then, Newton's identities (or the Newton-Girard formulae) provide a
set of relations between the traces,
\begin{equation}
t_1 - c_{n-1} = 0, \quad t_k - c_{n-1} t_{k-1} + \ldots - c_{n-k+1} t_1 + k \cdot c_{n-k} =
0,\quad k=2,3,\ldots, n \, ,
\end{equation}
which can be solved recursively. The solution can conveniently be
written down in closed form as
\begin{equation}
c_{n-k} = \frac{1}{k!} \det \left(
\begin{array}{cccccc}
t_1 & 1   & 0  & 0 & \cdots & 0 \\
t_2 & t_1 & 2  & 0 &  \cdots      & 0 \\
t_3 & t_2 & t_1& 3 &  \cdots      & 0 \\
 \vdots   &  \vdots & \vdots &  \vdots & \ddots & \vdots\\
t_{k-1} & t_{k-2} & t_{k-3} & t_{k-4} & \cdots & k-1 \\
t_{k} & t_{k-1} & t_{k-2} & t_{k-3} & \cdots & t_1 
\end{array}
\right) \, 
\end{equation}
and the generalisation to periodic b.c.~is again straighforward.

\subsection{Coefficients in terms of minors}
\label{app:coefficients from minors}
Instead of computing the traces of the matrices ${\cal T}, {\cal T}^2,
{\cal T}^3, \ldots, {\cal T}^n$ we now present an alternative method
for determining the coefficients of the characteristic polynomial
which is more interesting from the point of view of the transfer
matrix construction discussed in section \ref{sec:fermion
  sectors}. The method involves the expansion of determinants of order
1 to $n$ \cite{Pennisi:1987}. In order to determine the coefficients
$c_k$ of $x^k$ in $g(x)$ it is useful to separate the occurrences of
$x$ by introducing
\begin{equation}
f(x_1,x_2,\ldots,x_n) = \det\left({\cal T} +
  \text{diag}(x_1,x_2,\ldots\,x_n)\right) \, .
\end{equation}
One then has $g(x) = f(x,x,\ldots,x)$ and $c_k$ is the sum of the
coefficients of the terms with total degree $k$ in
$f(x_1,x_2,\ldots,x_n)$. Since $f(x_1,x_2,\ldots,x_n)$ is of degree 1
in each $x_i$, it is straightforward to express the coefficient in
terms of derivatives w.r.t.~$x_i$'s,
\begin{equation}
c_k = \left. \sum_{1\leq i_1<\cdots<i_k \leq n} \frac{\partial^k}{\partial
  x_{i_1} \partial x_{i_2} \cdots\partial x_{i_k} }
f(x_1,x_2,\ldots,x_n) \right|_{x_1=x_2=\ldots=x_n=0} 
\end{equation}
where $0 \leq k \leq n$. As a consequence the coefficients are now
expressed explicitly in terms of the matrix elements of ${\cal
  T}$. Denoting them by $t_{ij}$ it turns out that
\begin{equation}
\label{eq:coefficient c_k}
c_k = \sum_{1\leq i_1<\cdots<i_k \leq n} \frac{\partial^k}{\partial
  t_{i_1 i_1} \partial t_{i_2 i_2} \cdots\partial t_{i_k i_k} } \det
{\cal T} \, .
\end{equation}

This can be seen most easily by suppressing the dependence of $\det
{\cal T}$ on the off-diagonal elements $t_{ij}, i\neq j$ and define
$D$ as a function of the $n$ variables $t_{11}, t_{22}, \ldots,
t_{nn}$,
\begin{equation}
\label{eq:DeqDtilde}
D(t_{11},t_{22},\ldots,t_{nn}) \equiv \det {\cal T} \, ,
\end{equation}
and hence
\begin{equation}
f(x_1,x_2,\ldots,x_n) = D(t_{11}+x_1,t_{22}+x_2,\ldots,t_{nn}+x_n) \, .
\end{equation}
It is then immediately clear that 
\begin{equation}
\frac{\partial^k f}{\partial x_{i_1} \partial x_{i_2} \ldots \partial
  x_{i_k}} 
=
\frac{\partial^k  D(t_{11}+x_1,t_{22}+x_2,\ldots,t_{nn}+x_n)}{\partial  t_{i_1 i_1} \partial t_{i_2 i_2} \cdots\partial t_{i_k i_k} } \, .
\end{equation}
from which eq.(\ref{eq:coefficient c_k}) follows via
eq.(\ref{eq:DeqDtilde}).

On the other hand the rules for the Laplace expansion of a determinant
by a row or a column indicate that $\partial \det {\cal T}/\partial
t_{ij}$ is the $(i,j)$-cofactor of ${\cal T}$, or in fact the
$(i,i)$-minor when $i=j$. Therefore, the partial derivatives in
eq.(\ref{eq:coefficient c_k}) are simply the subdeterminants of ${\cal
  T}$ resulting from crossing out the rows and columns numbered by
$i_1,i_2,\ldots,i_k$, i.e., the principal minors of ${\cal T}$ of
order $k$.

Denoting the sum of principal minors of order $k$ of ${\cal T}$ by
$E_k({\cal T})$ and keeping in mind that $\det {\cal D}_k = c_k$ one
finds by comparison with eq.(\ref{eq:Dk symmetric}) that
\begin{equation}
S_{n-k}({\cal T}) = E_k({\cal T})
\end{equation}
for each $k=1,\ldots, n$, which is a known identity in matrix analysis
from linear algebra, see e.g.~\cite{Horn:2013}.

Comparing these results with the ones derived in section
\ref{sec:fermion sectors} we immediately notice that the trace over
the states of the transfer matrix is represented in
eq.(\ref{eq:coefficient c_k}) by the sum
$\sum_{i_1<i_2<\ldots<i_k}$. The number of summands here is $\left( n
  \atop k \right)$ and indeed equal to the number of states in the
sector with $n_f=k$. Furthermore, the principal subdeterminants
(minors) in eq.(\ref{eq:coefficient c_k}) correspond to the diagonal
elements of the product of transfer matrices in the given sector.

\section{Equivalence of canonical determinants}
\label{app:equivalence canonical determinants}
Here we show that the canonical determinants obtained in the fermion
loop approach, cf.~eq.(\ref{eq:generic canonical det}), are equal to
the ones using the fermion matrix reduction, cf.~eq.(\ref{eq:Dk
  symmetric}).

Following the notation introduced in section \ref{sec:generic sector},
for two index sets $A$ and $B$ of size $n_f$ the transfer matrix
$T_{n_f}^\Phi$ in eq.(\ref{eq:TPhi}) is the transposed matrix of
cofactors of $\Phi$ of order $n_f$ and is denoted by
\begin{equation}
\left(T^\Phi_{n_f}\right)_{AB} = C_{\, \backslash \hspace{-0.2cm} B \, \backslash
  \hspace{-0.2cm}A}(\Phi)  \, ,
\end{equation}
while the transfer matrix $T_{n_f}^W$ in eq.(\ref{eq:TW}) is the
matrix of complementary minors denoted by
\begin{equation}
\left(T^W_{n_f}\right)_{AB} = M_{AB}(W)  \, .
\end{equation}
Now we note that the complementary minor matrix $M_{AB}(W)$ is related
to the minor matrix of the inverse $M_{\,\backslash \hspace{-0.2cm} A
  \, \backslash \hspace{-0.2cm}B}(W^{-1})$ by
\begin{equation}
M_{\,\backslash \hspace{-0.2cm} A  \, \backslash
  \hspace{-0.2cm}B}(W^{-1}) = (-1)^{p(A,B)}
\frac{M_{BA}(W)}{\det W} 
\end{equation}
where $p(A,B) = \sum_{i\in A} i + \sum_{j\in B} j$. Up to the
determinant, the r.h.s.~is the higher order generalisation of the
adjugate (or classical adjoint) of $W$, i.e.~$\text{Adj}_{AB}(W)$. (To
order 1 the adjugate is just the transposed complementary cofactor
matrix.) Hence, with $\det W = 1$, $W^{-1} = W^\dagger=W^T$ and
$M_{AB}(W) = M_{BA}(W^T)$ we have
\begin{equation}
C_{\,\backslash \hspace{-0.2cm} B  \, \backslash
  \hspace{-0.2cm}A}(W^\dagger) = C_{\,\backslash \hspace{-0.2cm} A  \, \backslash
  \hspace{-0.2cm}B}(W) = M_{AB}(W) \, ,
\end{equation}
i.e., the transfer matrix $T_{n_f}^W$ can be expressed as a cofactor
matrix instead of a complementary minor matrix.

Next, we note that the cofactor matrix $C$ and the corresponding minor
matrix $M$ are related by modifying the sign of each element according
to $C_{AB} = (-1)^{p(A,B)} M_{AB}$.  The sign change can be achieved
by a similarity transformation with the matrix
$S_{AB}=(-1)^{\sum_{i\in A} i} \, \delta_{AB}$, i.e., $C = S^{-1}
\cdot M \cdot S$. Therefore a product of cofactor matrices becomes a
product of minor matrices under a trace, and so we can eventually
write
\begin{align}
\det {\cal D}_{n_f} &= \Tr \prod_{t=0}^{L_t-1}
\left[T_{n_f}^\Phi(t) \cdot T^W_{n_f}\right] \\
 &= \Tr \prod_{t=0}^{L_t-1} \left[C(\Phi(t))^T \cdot C(W) \right]\\
 &= \Tr \prod_{t=0}^{L_t-1} \left[M(\Phi(t)) \cdot M(W)
 \right] \, .
\label{eq:product of minors}
\end{align}
Note that we have made use of the fact that $C(\Phi)^\dagger =
C(\Phi)$ since $\Phi^\dagger = \Phi$.

We can now employ the Cauchy-Binet formula which states in its
symmetric form that given the $n\times n$ matrices $P,Q$ with $R=P Q$
and two index sets $A,B$ of size $1\leq k \leq n$ the $(AB)$-minor of
$R$ is
\begin{equation}
\det R^{\,\backslash\hspace{-0.2cm} A \,\backslash \hspace{-0.2cm}B}
= \sum_D \det P^{\,\backslash \hspace{-0.2cm} A  \, \backslash  \hspace{-0.2cm}D}
\det Q^{\,\backslash \hspace{-0.2cm} D  \, \backslash  \hspace{-0.2cm}B}
\end{equation}
where the sum is taken over all index sets $D$ of size $k$. From the
formula it follows that for the matrices of minors (and similarly for
the matrices of cofactors) one has
\begin{equation}
M(P Q) = M(P) M(Q) 
\end{equation}
and consequently from eq.(\ref{eq:product of minors})
\begin{align}
\det {\cal D}_{n_f} &=
 \Tr \prod_{t=0}^{L_t-1} \left[M(\Phi(t)) \cdot M(W)
 \right] \\
  &= \Tr M\left( \prod_{t=0}^{L_t-1} \left[\Phi(t) W\right]
  \right) \\
  & = \Tr M({\cal T}) \, .
\end{align}

Finally, the trace sums over the $\left( n_f^\text{max} \atop n_f
\right)$ diagonal elements of the minor matrix which are just the
principal minors,
\begin{equation}
\det {\cal D}_{n_f} = \sum_B \det {\cal T}^{\,\backslash \hspace{-0.2cm} B  \, \backslash
  \hspace{-0.2cm}B} \equiv E_{n_f}({\cal T})\, .
\end{equation}
Recalling from linear algebra \cite{Horn:2013} the fact that the sum
of all principal minors of order $n_f$ of a matrix is equal to the
$(n_f^\text{max} - n_f)^{th}$ symmetric function of its eigenvalues,
i.e.~$E_{n_f}({\cal T}) = S_{n_f^\text{max} - n_f}({\cal T})$,
eventually proves the equivalence between $\det {\cal D}_{n_f}$ from
the fermion loop formulation in eq.(\ref{eq:Dk minors}) and from the
fermion matrix reduction in eq.(\ref{eq:Dk symmetric}).

\end{appendix}

\bibliographystyle{JHEP}
\bibliography{lfoSYMQM}

\end{document}